%% file: ms.tex
\documentclass[final]{pasj00}

\Received{$\langle$reception date$\rangle$}
\Accepted{$\langle$acception date$\rangle$}
\Published{$\langle$publication date$\rangle$}
\SetRunningHead{Astronomical Society of Japan}{Usage of \texttt{pasj00.cls}}

\usepackage{times}

\begin{document}

\title{Two Types of Magnetohydrodynamic Sheath Jets}
\author{Osamu Kaburaki%
\thanks{Present Address is 
Oh-uchi Yata, Yamaguchi-shi, Yamaguchi 753-0215}}
\affil{Domain for Fundamental Sciences, Graduate School of Science and Engineering, 
Yamaguchi University, Yamaguchi 753--8512}
\email{kaburaki@sci.yamaguchi-u.ac.jp}


\maketitle

\begin{abstract}
Recent observations of astrophysical jets emanating from various galactic nuclei strongly 
suggest that a double layered structure, or a spine-sheath structure, is likely to be their 
common feature. 
We propose that such a sheath jet structure can be formed magnetohydrodynamically 
within a valley of the magnetic pressures, which is formed between the peaks due to the poloidal 
and toroidal components, with the centrifugal force acting on the rotating sheath plasma 
is balanced by the hoop stress of the toroidal field. 
The poloidal field concentrated near the polar axis is maintained by a converging plasma flow 
toward the jet region, and the toroidal field is developed outside the jet cone owing to the 
poloidal current circulating through the jet. Under such situations, the set of magnetohydrodynamic 
(MHD) equations allows two main types of solutions, at least, in the region far from the footpoint.
The first type solution describes the jets of marginally bound nature. This type is realized 
when the jet temperature decreases like viral one, and neither the pressure-gradient nor 
the MHD forces, which are both determined consistently, cannot completely overcome the gravity 
even at infinity. The second type is realized under an isothermal situation, and the gravity 
is cancelled exactly by the pressure-gradient force. Hence, the jets of this type are accelerated 
purely by the MHD force. It is suggested also that these two types correspond, respectively, 
to the jets from type I and II radio galaxies in the Fanaroff-Riley classification. 

\end{abstract}

\input{pt1.tex}

\input{pt2.tex}

\end{document}

%% file: pt1.tex

\section{Introduction}

In his penetrating review paper on the active galactic nuclei (AGN), \citet{Lhy99} stresses the 
importance of the dichotomy of radio loud and quiet AGN and suggests the essential difference 
between the respective central engines. In this context, our interest in the present paper 
is substantially restricted to the radio loud AGN and their power-down version, the low luminosity 
active galactic nuclei (LLAGN). 

It is well known (see, e.g., \cite{Ury95,DPn05}) that there are two kinds of astrophysical 
jets associated with the type I and II radio galaxies in the Fanaroff-Riley (or FR) classification 
\citep{FR74}. The radio emissions of the low-luminosity FR\ I objects peak near the nuclei and often 
show symmetric two-sided jets. The collimations of the jets are not so strong and the radial 
velocities show considerable deceleration at large distances. These objects are found in rich 
cluster environments. On the other hand, the emissions from the high-luminosity FR\ II objects have 
radio lobes with prominent hot spots and bright outer edges. The collimations of the jets are very 
sharp and strong asymmetries often exist between the jets and counter jets, with the latter being 
undetectable in many cases. The radial velocities do not seem to be decelerated until the end 
points are reached. These objects exist in more isolated environments than FR\ I's. 

Another feature we have to keep in mind is the double-layered structure, or spine-sheath structure, 
of the jets, which is now believed to be a common feature to the FR\ I and II jets \citep{Giv99,Giv01}. 
The limb-brightened structure observed in the jets of both types can be well interpreted by the 
presence of velocity shear associated with the double-layered structure \citep{Gir04}, in which 
a spine region of highly relativistic speed is enclosed by a mildly relativistic or non-relativistic 
sheath jet. Owing to the relativistic beaming effect, the radiation from the spine becomes highly 
concentrated within a small angle around the jet direction, and hence becomes weak when seen from 
other directions. The possibility that the plasma in the spine region may be the electron-positron 
pair plasma has long been suggested in the literature (see, e.g., \cite{BSR87,Plt04}).  
 
Historically, the problem of jet formation has been discussed viably in the framework of the 
theory of force-free or ideal magnetohydrodynamic (MHD) magnetospheres (e.g., \cite{Sty02,Plt04}) 
originally developed in relation to the discussions of the pulsar magnetospheres. Although such 
treatments may be mathematically sophisticated and hence beautiful, it seems unfortunately too 
formal to be able to quickly solve many practical problems associated with the actual astrophysical 
jets. Therefore, we adopt here a much more practical approach to this problem. The essence of the jet 
formation and acceleration is in their non-force-free nature and non-ideal MHD aspects of the 
electrodynamic current systems. 

As stressed in our previous works (e.g., \cite{Kab00,Kab07}), the jets can be formed by pinching 
and accelerating effects of the Amp\'ere force acting on the current carrying plasma in the polar 
regions. The current flowing through the polar jet forms a returning portion of the poloidally 
circulating current system (see, figure 1). This current is originally driven by the electromotive 
force caused in the accretion disc by the rotational motion of the disc plasma in a global, poloidal 
magnetic field. The mechanism of jet formation is thus directly coupled to the accretion process 
taking place in the same activity site. 
Although there are several streams of investigations that attribute the formation of jets or outflows 
to the action of magnetized accretion discs (e.g., \cite{BP82}), no satisfactory theory from 
the view point of the current closure seems to exist so far in the literature. The importance of 
viewing the jet as a part of a globally circulating current system has been emphasized from early 
times by \citet{Ben78}, and remarkable stability natures of such current-carrying systems have also 
been discussed repeatedly by the same author (e.g., \cite{Ben06}). 

Recent observations have also revealed that both types of jets in FR classification are likely to 
be enclosed within the X-ray cavities, as far as they are young. The cavities are probably 
maintained by the magnetic pressure of the toroidal field generated by the global current system 
circulating the disc-jet compounds \citep{Kab07}. It is interesting to note that the shapes of 
such cavities may be different reflecting the different types of the current systems enclosed 
therein (see, figure 1 and also the discussion in the final section). 
It is theoretically probable that if a jet is being driven by the accretion disc the resulting 
cavity would take an hour-glass shape, whereas the cavity would become of spindle shape 
if the accretion has been switched off. 
Observationally, the cavities of FR\ I jets seem to be of hour-glass shape with narrow necks near 
the equator (e.g., for Cen A, see figure 2 and 3 of \cite{Krf02}, and figure 4 of \cite{Krf07}; 
for M87, figure 6 of \cite{For07} 
). On the other hand, 
the examples of the spindle shape are fairly rare. The only one plausible example hitherto known 
to the author is Cyg A (see figure 1 of \cite{WSY06}), which belongs to FR\ II type.

\begin{figure}


 \begin{center}
  \FigureFile(70mm, 85mm){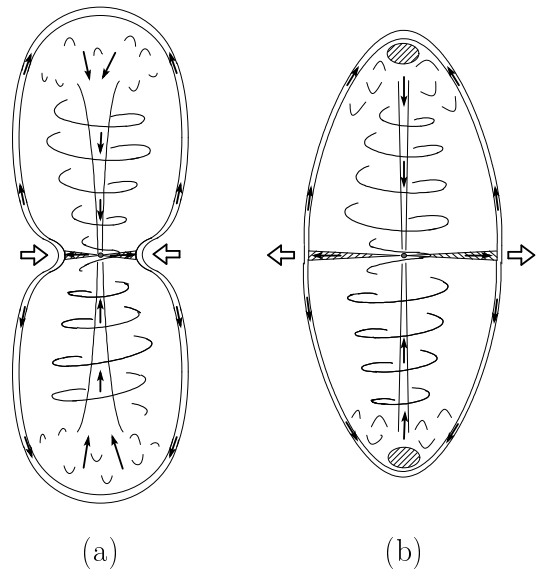}
 \end{center}
 \caption{
A theoretically suggested classification scheme of the X-ray jet cavities. The paths of the globally 
circulating electric current are shown by black arrows for the case of an upward directed seed field. 
Also, the directions of the radial component of equatorial flows are indicated by thick 
white arrows. 
(a) Hour-glass shaped jet cavity, typical of accretion-driven jets. The appearance of a narrow neck 
near the equatorial plane is caused by the presence of inward flow toward the accretion disc. 
The electric current is being driven by the accretion disc around the central black hole. This 
shape seems to be a common feature of FR\ I jets. 
(b) Spindle shaped jet cavity, associated with a switched-off current system. The equatorial plasma 
disc is not driving the current system because of the lack of matter supply, and instead, the current 
is maintained by the huge self-inductance of the system. The appearance of the inductive electric 
field causes an outward flow in the equatorial disc (see the discussion in section 7). 
}
\end{figure}

Our aim in the present paper is to construct an analytic model of conical MHD jets with small opening 
angles, which can reproduce the main properties of the jets described above, at least, in the regions 
well beyond their launching site. The method of analysis adopted here is analogous to that in our 
earlier work \citep{KabI87} addressed to the jets in star forming regions. The essential difference 
between the situations considered here and there is in the assumed external magnetic fields. Instead 
of a stellar dipole field in the former paper, we assume here a uniform field perpendicular to the 
equatorial plasma disc, in the context of galactic nuclei. The effects of relativity are not included 
for simplicity, since the jet velocities in the MHD sheaths are expected to be at most mildly 
relativistic. Preliminary results have been reported in \citet{Kab04}, though it contains the solution 
of only one out of two types obtained in the present paper.  

In section 2 we first describe the global configuration considered in the present paper, and then 
specify the angular dependences of relevant physical quantities in spherical polar coordinates. 
With these expressions, we discuss in section 3 Ohm's law and the equation of motion, which are 
essential to determine the jet velocities. In order to obtain the solutions to this set of equations, 
we proceed in  section 4 from the static force balance to a plausible case of dynamic balance. In the 
following two sections, two types of solutions are derived and their properties are discussed 
in detail. Finally in section 7, the results are summarized and then a possible interpretation 
of these solutions is discussed in the light of the FR dichotomy of the galactic radio jets.   

\section{Separation of Variables}

When the processes taking place in the central accretion disks are explicitly taken into account, 
the mechanisms of jet formation are expected to be as follows. 
In terms of the circuit theory, the accretion disc acts as a DC current generator and drives 
poloidally circulating current system (for more details, see section 2 of \cite{Kab07} and 
figures therein). This is the physical cause of the toroidal magnetic component that is added 
as a twist to the originally vertical seed field. 
In the case of an upward polarity of the seed field, the radial current is driven outward 
in the disc, and dominant portion of this current is expected to close after circulating remote 
regions along the thin boundary layer separating the jet cavity (cocoon or hour glass) from 
surroundings. This means that the twisted field lines are confined exclusively within the cavity. 
After passing through the tops of the cavity wall, the currents return to the inner edge 
of the accretion disc along the polar axes. 

Under such circumstances, a polar jet can be formed from plasma in the polar regions, because 
the Amp\`ere force exerted by the toroidal magnetic field on the return current always has 
the necessary components both for radial acceleration and for collimation, as shown in figure 1 
of \citet{Kab07}. This fact implies that there may always be a bipolar jet associated with 
such an accretion disc in a large-scale magnetic field, even if it might be strong or rather 
weak depending on the circumstances. 

In an axisymmetric situation, the only relevant variables in spherical polar coordinates 
($r$, $\theta$, $\varphi$) are the radial distance from the center $r$ and the polar angle 
$\theta$. It is often convenient to introduce their normalized versions, $z\equiv r/r_{\rm J}$ and 
$\eta\equiv \theta/\Theta$, where $r_{\rm J}$ is the footpoint radius and $\Theta$ is the 
half-opening angle, respectively, of a conical jet. The angular variable $\eta$ introduced here 
should not be confused with that used in our previous papers dealing with the problems of disc 
accretion (\cite{Kab00, Kab01, Kab07}; in these papers, $\eta$ stands for $(\theta-\pi/2)/\Delta$, 
with $\Delta$ being the half-opening angle of an accretion disc). The magnetic field is generally 
written as ${\bf B_0}+{\bf b}$, where ${\bf B_0}$ is the seed field and ${\bf b}$ is the deformation 
due to the action of plasma motion. However, the latter is sufficiently large (i.e., 
$\vert{\bf b}\vert\gg\vert{\bf B_0}\vert$) in the accretion discs and jets. The influence of the 
seed field is taken into account only at the outer edge of an accretion disc (see \cite{Kab00}). 

Jets are assumed to be narrow (i.e., $\Theta\ll1$), and $\Theta$ is regarded as a smallness 
parameter in ordering the physical quantities. 
Our main interest is not in obtaining a strict solution, but in demonstrating the existence of 
a physically reasonable configuration that satisfies the set of equations, at least in an approximate 
sense. We therefore expect that the solution to the set of resistive-MHD equations can be written 
in a variable-separated form. 
The concrete form of each physical quantity is determined here by taking its symmetry property 
into account, but we do not insist that such a choice is the unique possibility. 

We adopt the following functional forms for the three components of the magnetic field: 
\begin{equation}
     b_r(z, \eta) =\ \tilde{b}_r(z)\ \mbox{sech}^2\eta, 
  \label{eqn:br}
\end{equation}
\begin{equation}
     b_{\theta}(z, \eta) = -\Theta\tilde{b}_{\theta}(z) f(\eta),  
 \label{eqn:bth}
\end{equation}
\begin{equation}
     b_{\varphi}(z, \eta) = -\tilde{b}_{\varphi}(z)\tanh\eta. 
\end{equation}
Here we have introduced the functions carrying tildes over them to represent the radius-dependent 
parts of the corresponding quantities. Although they are written as functions of normalized radius 
$z$, but they can also be regarded as functions of $r$ (i.e., radius itself) whenever it is necessary.  
Hereafter, we discuss only one side of a bipolar-jet system, and hence the values of $\theta$ are 
restricted to the range $0\leq\theta\leq\pi/2$ whereas the azimuthal angle varies in the whole range 
$0\leq\varphi\leq 2\pi$. 

The $\eta$-dependence of $b_r$ is assumed to represent a localized structure within the cone of 
$\theta\sim\Theta$, which is expected to occur in this component owing to the balance between 
the sweeping up effect of the accreting flow at the footpoint of a jet and the outward diffusion 
caused by the presence of a finite electrical resistivity. 
The appearance of the smallness parameter $\Theta$ in $b_{\theta}$ indicates that this quantity is 
of order $O(\Theta)$, and the concrete form of the function $f(\eta)$ is specified below from the 
requirement of flux conservation. The presence of $\tanh\eta$ in $b_{\varphi}$ guarantees that 
$b_{\varphi}$ should vanish on the polar axis and also remain finite outside the jet cone. 

The sign convention on the right-hand sides of the above expressions is adopted for an upward seed 
field whose $r$-component on the polar axis is positive. When the seed field is directed downward, 
$B_0$ in the resulting solutions should be regarded as having minus sign, and any other inclined case 
is not considered here. 
Although the seed field is amplified strongly by the inward motion in an accretion disc, its 
polarity remains the same. The rotational motion in the disc causes a large negative 
$\varphi$-component, which is assumed to be absent in the seed field. Further, the negative sign 
in the $\theta$-component means that a converging magnetic field in the jet structure is 
taken as standard. 

For the expressions (\ref{eqn:br}) and (\ref{eqn:bth}), the law of magnetic flux conservation 
${\bf \nabla}\cdot{\bf b} = 0$ yields 
\begin{equation}
  \frac{1}{r^2}\frac{d}{dr}(r^2\tilde{b}_r)\ \mbox{sech}^2\eta
   -\frac{\Theta\tilde{b}_{\theta}}{r\sin\theta}\frac{d}{d\theta}
    \left(f\sin\theta\right) = 0. 
 \label{eqn:divb}
\end{equation}
As far as we remain within the jet cone (i.e., $\theta\leq\Theta$), the separation of 
variable is attained approximately by the choice 
\begin{equation}
  f(\eta) = \tanh\eta - \frac{\Theta}{\sin\theta}\ln\left(\mbox{cosh}\ \eta\right), 
 \label{eqn:f}
\end{equation}
where the second term on the right-hand side is of the first order in $\Theta$ 
since $\ln(\mbox{cosh}\ \eta)/\sin\theta$ remains finite, even for very small $\theta$. 
Then the conservation equation reduces to 
\begin{equation}
   \frac{\tilde{b}_\theta}{\tilde{b}_r} 
  = r\frac{d}{dr}\ln(r^2\tilde{b}_r). 
 \label{eqn:fcons}
\end{equation}
In most cases, however, we need only the expression of $f(\eta)$ that has the accuracy to the 
leading (i.e., zero-th) order in $\Theta$ as far as we remain in the jet-cone region stated above. 
In such cases, we can replace (\ref{eqn:f}) by its approximate version, 
\begin{equation}
  f(\eta) \simeq \tanh\eta .
 \label{eqn:aprf}
\end{equation}
This procedure corresponds to regarding $\sin\theta$ in the $\eta$-derivative in equation 
(\ref{eqn:divb}) as a slowly varying quantity (i.e., a constant) compared with a rapidly varying 
function $\tanh\eta$. Then, we can reproduce equation (\ref{eqn:fcons}) within this approximation. 
Exceptional cases in which the accurate expression (\ref{eqn:f}) is needed are discussed in the next 
section. 

Given the above expressions for the magnetic field components, we can derive the forms of the current 
density from Amp\`ere's law 
\begin{equation}
  {\bf j} = \frac{c}{4\pi}{\bf \nabla}\times{\bf b}, 
\end{equation}
where $c$ is the light velocity in a vacuum. 
Their components are 
\begin{eqnarray}
 j_r(z, \eta) &=& \frac{c}{4\pi r\sin\theta}\ \frac{\partial}{\partial\theta}
   \left( \sin\theta\ b_{\varphi} \right) \nonumber \\
  & \simeq & -\Theta^{-1}\tilde{j}_r(z)\ \mbox{sech}^2\eta, 
  \label{eqn:jr}
\end{eqnarray}
\begin{equation}
 j_{\theta}(z, \eta) 
   = -\frac{c}{4\pi r}\frac{\partial}{\partial r}\left( rb_{\varphi} \right) 
   = \tilde{j}_{\theta}(z)\tanh\eta, 
\end{equation}
\begin{eqnarray}
 j_{\varphi}(z, \eta) 
  &=& \frac{c}{4\pi r}\left\{ \frac{\partial}{\partial r}(rb_{\theta})
    -\frac{\partial b_r}{\partial\theta} \right\}  \nonumber \\
  & \simeq & \ \Theta^{-1}\tilde{j}_{\varphi}(z)\ \mbox{sech}^2\eta\ \tanh\eta, 
\label{eqn:jphi}
\end{eqnarray}
where 
\begin{equation}
 \tilde{j}_r(z) 
   = \frac{c}{4\pi r_{\rm J}}\ \frac{\tilde{b}_{\varphi}(z)}{z}, 
 \label{eqn:tjr}
\end{equation}
\begin{equation}
 \tilde{j}_{\theta}(z) = \frac{c}{4\pi r_{\rm J}}\ 
  \frac{1}{z}\frac{d}{dz}[z\tilde{b}_{\varphi}(z)], 
 \label{eqn:tjth}
\end{equation}
\begin{equation}
 \tilde{j}_{\varphi}(z) = \frac{c}{2\pi r_{\rm J}}\ \frac{\tilde{b}_r(z)}{z}. 
 \label{eqn:tjphi}
\end{equation}

The factor $\Theta^{-1}$ that expresses the order of magnitude of the relevant quantity has 
been made explicit in equations (\ref{eqn:jr}) and (\ref{eqn:jphi}). In performing 
the $\eta$-derivative in equation (\ref{eqn:jr}), $\sin\theta$ has been regarded also as 
a constant as noted above. In deriving the last expression of equation (\ref{eqn:jphi}), 
the first term in the curly brackets has been omitted compared with the second term, because 
the former is of order $\sim O(\Theta)$ while the latter is $\sim O(\Theta^{-1})$ owing to 
the presence of the derivative $\partial/\partial\theta=\Theta^{-1}\partial/\partial\eta$. 
The negative sign in front of $\tilde{j}_r$ means that the incoming current is a standard 
when the seed field is directed upward. 
It is easy to confirm that the above expressions actually satisfy ${\bf \nabla}\cdot{\bf j} = 0$. 

The functional forms of the velocity components are introduced as follows: 
\begin{equation}
     v_r(z, \eta) =\ \tilde{v}_r(z)\ \mbox{sech}^2\eta, 
\end{equation}
\begin{equation}
     v_{\theta}(z, \eta) 
  = -\Theta\tilde{v}_{\theta}(z) h(\eta)
  \label{eqn:vth}
\end{equation}
\begin{equation}
     v_{\varphi}(z, \eta) 
    = \tilde{v}_{\varphi}(z) \tanh\eta.
\end{equation} 
Again, the hyperbolic functions and $h(\eta)$ represent the existence of a localized structure, 
but the sign convention here is somewhat different from the case of magnetic field. The jet 
proceeds outward, of course, and is expected to rotate in the same direction as the accretion disc 
and to converge toward the polar axis. When $\tilde{v}_r(z)$, $\tilde{v}_{\theta}(z)$ and 
$\tilde{v}_{\varphi}(z)$ are regarded as quantities of order unity, only $v_{\theta}(z, \eta)$ 
is of order $\sim O(\Theta)$ owing to the presence of a factor $\Theta$. Although $v_{\varphi}$ 
remains finite even outside the jet cone, the plasma density vanishes there (see below). 

Analogously to the case of the flux conservation, the equation of mass continuity 
${\bf \nabla}\cdot(\rho{\bf v}) = 0$ can be reduced to the form 
\begin{equation}
   \frac{\tilde{v}_\theta}{\tilde{v}_r} 
  = r\frac{d}{dr}\ln(r^2\tilde{\rho}\tilde{v}_r), 
 \label{eqn:mcons}
\end{equation} 
by fixing as 
\begin{equation}
  h(\eta)=f(\eta). 
\end{equation}
Of course, however, some difference from the former case arises according to the appearance of 
an extra function $\rho$ in the continuity equation. The angular dependence of the matter 
density is assumed to be 
\begin{equation}
     \rho(z, \eta) =\ \tilde{\rho}(z)\ 
    \mbox{sech}^2\eta \tanh^2\eta, 
\end{equation}
reflecting a hollow-cone structure suggested by observations. Then, the density maximum is 
attained where $d\rho/d\eta=0$, i.e. where $\mbox{sech}^2\eta=\tanh^2\eta$, and we identify this 
surface of maximum density as the middle surface of a hollow-cone jet (i.e., $\theta\sim\Theta$). 
In the above derivation of equation (\ref{eqn:mcons}), therefore, the term that is proportional 
to $d(\mbox{sech}^2\eta\tanh^2\eta)/d\eta$ has been omitted. 

The remaining thermodynamic quantities are assumed to take the following forms:
\begin{equation}
     T(z, \eta) = \tilde{T}(z), 
\end{equation}
\begin{equation}
     p(z, \eta) =\ \tilde{p}(z)\ 
    \mbox{sech}^2\eta \tanh^2\eta. 
\end{equation}
Here, the variation of the temperature $T$ across a jet is ignored since the jet is very thin.
The $\eta$-dependence of the pressure represents also a hollow-cone structure. 
We should emphasize again that all of the functional forms introduced above are, by no means, 
insisted as a unique possibility for each relevant quantity, but that we are searching only 
for a plausible example. 

Then, the ideal gas law $p = K\rho T$ is separated completely to yield 
\begin{equation}
    \tilde{p} = K\tilde{\rho}\tilde{T},
 \label{eqn:gasp}  
\end{equation}
where $K=R/\mu$, with $R$ and $\mu$ being the gas constant and the mean molecular weight 
of the jet plasma, respectively.  

\section{Ohm's Law and Equation of Motion}

The remaining fundamental equations are Ohm's law and equation of motion. Although the electric 
field is usually eliminated from the scheme of MHD, especially in its ideal version, we treat it 
explicitly here. This is because it terns out that Ohm's law, rather than the equation of motion, 
plays an essential role in determining the radial velocity of a jet, at least in the region far 
from the footpoint. It turns out unfortunately that some components of Ohm's law and the equation 
of motion (EOM) cannot be variable-separated completely. In such cases, we approximately replace 
the remaining angluar dependences by their representative values evaluated at the middle surface 
of a jet cone, $\theta\sim\Theta$, where most of the jet matter is concentrated. For example, 
whenever the dependences on ${\rm sech}^2\eta$ and $\tanh^2\eta$ appear, we evaluate their 
representative values as $\langle\mbox{sech}^2\eta\rangle=\langle\tanh^2\eta\rangle = 1/2$, 
because these satisfy the condition for the maximum density, $\mbox{sech}^2\eta=\tanh^2\eta$, and 
the identity $\mbox{sech}^2\eta+\tanh^2\eta=1$, simultaneously. 

When the system of current carrying jet satisfies a strict stationarity condition, the electric 
field should be irrotational and hence can be written as the gradient of a scalar potential 
$U(z, \eta)$, i.e., ${\bf E}=-{\bf \nabla}\ U$. Reflecting the localized structure associated 
with the jet, we assume for the potential as 
\begin{equation}
  U(z, \eta) = \Theta\tilde{U}(z)\ \mbox{sech}^2\eta,  
\end{equation}
and obtain 
\begin{equation}
  E_r(z, \eta) = -\frac{\partial U}{\partial r} 
   = -\Theta\tilde{E}_r(z)\ \mbox{sech}^2\eta, 
 \label{eqn:Er}
\end{equation}
\begin{equation}
  E_{\theta}(z, \eta) = -\frac{1}{r}\frac{\partial U}{\partial\theta} 
   = \tilde{E}_{\theta}(z)\ \mbox{sech}^2\eta\ \tanh\eta,  
\end{equation}
\begin{equation}
  E_{\varphi}(z, \eta) = -\frac{1}{r\sin\theta}\frac{\partial U}{\partial\varphi} 
      = 0, 
 \label{eqn:Ephi}
\end{equation}
where 
\begin{equation}
  \tilde{E}_r(z) = \frac{1}{r_{\rm J}}\frac{d\tilde{U}(z)}{dz},  
 \label{eqn:tEr}
\end{equation}
\begin{equation}
  \tilde{E}_{\theta}(z) = \frac{2}{r_{\rm J}}\frac{\tilde{U}(z)}{z}.  
 \label{eqn:tEth}
\end{equation}
The signs of $E_r$ and $E_{\theta}$ here have been chosen in accordance with those of $j_r$ 
and $j_{\theta}$, respectively. 

If the situation we are considering requires even a slow change of the magnetic field in time $t$, 
however, the electric field becomes non-irrotational according to Faraday's law,  
${\bf \nabla}\times{\bf E}=-(1/c)\ \partial{\bf b}/\partial t$. Assuming that even in such a case 
the functional forms of each component of ${\bf E}$ given above is not altered, we can show that 
the only non-vanishing value remains in its $\varphi$-component, 
\begin{equation}
  ({\bf \nabla}\times{\bf E})_{\varphi} 
   = \frac{1}{r}\left\{\frac{d}{dr}\left(r\tilde{E}_{\theta}\right) 
     -2\tilde{E}_r\right\}\ \mbox{sech}^2\eta\tanh\eta. 
 \label{eqn:rotE}
\end{equation}
Then, the $\varphi$-component of Faraday's law reduces to 
\begin{equation}
  \frac{\partial\tilde{b}_{\varphi}}{\partial t} 
   = -\frac{c}{r}\left\{\tilde{E}_r - \frac{1}{2}\frac{d}{dr}\left(r\tilde{E}_{\theta}\right)\right\}, 
 \label{eqn:dbdt}
\end{equation}
after the remaining $\eta$-dependence has been replaced by its representative value at the surface 
of the middle plane, $\theta\sim\Theta$. This equation can be used to judge whether the electric 
field obtained from Ohm's law satisfies a strict stationarity condition or not. Even if the electric 
field is non-irrotational, the rate of change for $b_{\varphi}$ is expected to be very small since 
the self-inductance of the whole current system is very large \citep{Ben06}. This point will be 
discussed after the explicit solutions are obtained. 

Ohm's law in its vector form is 
\begin{equation} 
  {\bf E} +\frac{1}{c}{\bf v}\times{\bf b} = \frac{\bf j}{\sigma},  
\end{equation}
where $\sigma$ is the electrical conductivity, and is assumed to be a constant for simplicity. 
This law yields 
\begin{equation} 
  \tilde{E}_r = \frac{1}{c}\ (\tilde{v}_{\theta}\tilde{b}_{\varphi} 
   + \tilde{v}_{\varphi}\tilde{b}_{\theta}) \ +\ \frac{\tilde{j}_r}{\sigma\Theta^2}, 
 \label{eqn:Ohr}
\end{equation}
\begin{equation} 
  \tilde{E}_{\theta} = -\frac{1}{c}\ (\tilde{v}_{\varphi}
   \tilde{b}_r + \tilde{v}_r\tilde{b}_{\varphi}), 
 \label{eqn:Ohth}
\end{equation}
\begin{equation} 
  \tilde{E}_{\varphi} = 0 = \frac{1}{c}(\tilde{v}_r\tilde{b}_{\theta} 
  - \tilde{v}_{\theta}\tilde{b}_r)
  \ +\ \frac{\tilde{j}_{\varphi}}{\sigma\Theta^2}, 
 \label{eqn:Ohphi}
\end{equation}
when the remaining $\eta$-dependences are replaced by their representative values. In deriving 
the above expressions, $\sigma$ has been regarded as a quantity of $O(\Theta^{-2})$. This is 
because the existence of the term $\tilde{j}_{\varphi}/\sigma$ in the $\varphi$-component equation 
is essential for taking the non-ideal MHD aspects into account. As a consequence, the term 
$\tilde{j}_{\theta}/\sigma$ has been omitted as small quantity of $O(\Theta^2)$ in equation 
(\ref{eqn:Ohth}). The origin of the resistivity may be classical or turbulent, but we do not need 
to specify it here because $\sigma$ appears in the final results only through the magnetic 
Reynolds number, which we may treat as a parameter. 

The MHD equation of motion in a stationary state is written as 
\begin{equation} 
    ({\bf v}\cdot{\bf \nabla}){\bf v} 
  + \frac{1}{\rho}\ {\bf \nabla}p 
  + \frac{1}{4\pi\rho}\ [{\bf b}\times({\bf \nabla}\times{\bf b})] 
  - {\bf g} = 0, 
\end{equation}
where ${\bf g}$ is the gravitational acceleration, and the contribution from the seed field 
${\bf B}_0$ has been neglected in the magnetic force term compared with that from the deformed part 
${\bf b}$. 

We first discuss the $\theta$-component of EMO. After writing out the full terms in this component, 
we keep only the leading order terms in the smallness parameter $\Theta$, i.e., the terms of 
$O(\Theta^{-1})$: 
\begin{equation}
  \left\{ v_{\varphi}^2-\frac{b_{\varphi}^2}{4\pi\rho} \right\}\mbox{cot}\theta 
  = \frac{1}{\rho}\frac{\partial}{\partial\theta}\left\{ p+\frac{1}{8\pi}(b_r^2+b_{\varphi}^2) 
   \right\}. 
 \label{eqn:thEOM}
\end{equation}
This equation states that if we expect the magnetic pressure confinement of the sheath plasma (i.e., 
the vanishing of the right-hand side), then the centrifugal force caused by the rotational motion 
of a jet plasma should be balanced by the hoop stress of the toroidal field lines. Although we do not 
insist that this is the only possibility to obtain a consistent set of solutions, we examine in this 
paper only the cases in which this condition is satisfied. It is easy to suppose that such a 
configuration would be realized in the magnetic fields developed in the galactic nuclear regions. 
Indeed, the radial component $b_r$ is expected to be packed into the spine region if a converging 
flow toward the polar axis is maintained, while the toroidal component $b_{\varphi}$ should remain 
weak near the polar axis as far as the poloidal current density is non-singular on the polar axis. 

Then, we obtain two separated equations: 
\begin{equation}
  v_{\varphi}^{\ 2} = \frac{b_{\varphi}^{\ 2}}{4\pi\rho}, 
  \quad  p+\frac{1}{8\pi}(b_r^2+b_{\varphi}^2) = \tilde{p}, 
 \label{eqn:EOMth}
\end{equation}
where $\tilde{p}$ is a function of $r$ only. 
The former equation yields 
\begin{equation}
  \tilde{v}_{\varphi}^{\ 2} 
  = \frac{\tilde{b}_{\varphi}^{\ 2}}{\pi\tilde{\rho}} 
 \label{eqn:v2phi}
\end{equation}
after substitution of the representative value for $\theta$, and the latter can be confirmed to 
hold if the condition 
\begin{equation} 
  \tilde{p} = \frac{\tilde{b}_r^{\ 2}}{8\pi}
  = \frac{\tilde{b}_{\varphi}^{\ 2}}{8\pi} 
 \label{eqn:bpxi}
\end{equation}
is satisfied. 

Similarly, retaining only the leading order terms in $\Theta$ in the $r$-component of EOM, we have 
\begin{eqnarray}
  \lefteqn{ v_r\frac{\partial v_r}{\partial r} 
   + \frac{v_{\theta}}{r}\frac{\partial v_r}{\partial\theta} 
   - \frac{v_{\varphi}^2}{r} + \frac{1}{\rho}\frac{\partial p}{\partial r} - g } \nonumber \\
   & & \qquad +\frac{1}{4\pi\rho r}\left\{ b_{\varphi}\frac{\partial}{\partial r}(rb_{\varphi})
   -b_{\theta}\frac{\partial b_r}{\partial\theta} \right\} = 0. 
\end{eqnarray}
In the above expression, the term $v_{\varphi}^2/r$ can be eliminated with the aid of the former 
one of equations (\ref{eqn:EOMth}). Although the gravity is usually neglected in the discussions 
of jet structures at large distances from the central object, here it is kept in the equation as a 
term of $O(1)$. Indeed, it turns out in the next section that the gravity plays an essential role 
in determining the structures of the actual jets, since the most fundamental thing that every jet 
should do is to overcome the gravity and to reach almost infinity. 

It should be noted that the accurate version of the function $f(\eta)$ in equation (\ref{eqn:f}) 
is conceived in the definition of the directional derivative $({\bf v}\cdot{\bf \nabla})$. Then, 
after substituting the functional forms of relevant quantities and evaluating the remaining 
$\eta$-dependences at $\theta\sim\Theta$, we obtain 
\begin{eqnarray} 
  \lefteqn{ \frac{\tilde{v}_r^{\ 2}}{4r} \left( r\frac{d\ln\tilde{v}_r}{dr}
    +2A\frac{\tilde{v}_{\theta}}{\tilde{v}_r} \right) } \nonumber \\
  & & \qquad + \frac{1}{\tilde{\rho}}\frac{d\tilde{p}}{dr} + \frac{GM}{r^2} 
    + \frac{\tilde{b}_{\varphi}^{\ 2}}{2\pi\tilde{\rho}r}
   \left( r\frac{d\ln\tilde{b}_{\varphi}}{dr} -D\frac{\tilde{b}_{\theta}}{\tilde{b}_r} \right) = 0, 
 \label{eqn:EOMr}
\end{eqnarray}
where $A$ and $D$ are newly introduced representative values, i.e., 
$A \equiv\langle f(\eta)\tanh\eta\rangle$ and $D \equiv\langle f(\eta)/\tanh\eta\rangle$. We can 
specify the range of these values as $0<A<1/2$ and $0<D<1$, since $f(\eta)<\tanh\eta$. However, 
as it is hard to specify these values more exactly, we treat them as parameters in this paper. 

In the $\varphi$-component of EOM, all the terms survive since they are all quantities of $O(1)$. 
After some manipulations, we can reach the form 
\begin{equation} 
  \left(v_r\frac{\partial}{\partial r}+\frac{v_{\theta}}{r}\frac{\partial}{\partial\theta}\right)l 
  - \frac{1}{4\pi\rho}
  \left(b_r\frac{\partial}{\partial r}+\frac{b_{\theta}}{r}\frac{\partial}{\partial\theta}\right)m 
  = 0, 
\end{equation}
where we have introduced the definitions 
\begin{eqnarray} 
  l \equiv rv_{\varphi}\sin\theta = \tilde{l} \sin\theta\tanh\eta, 
  \quad \tilde{l}\equiv r\tilde{v}_{\varphi}, \\
  m \equiv rb_{\varphi}\sin\theta = -\tilde{m} \sin\theta\tanh\eta, 
  \quad \tilde{m} \equiv r\tilde{b}_{\varphi}. 
\end{eqnarray}
Within the approximation that $d(\sin\theta\tanh\eta)/d\eta\simeq\sin\theta\ \mbox{sech}^2\eta$, 
we have 
\begin{equation} 
  \tilde{v}_r\tilde{l}\left( r\frac{d\ln\tilde{l}}{dr} 
  - D \frac{\tilde{v}_{\theta}}{\tilde{v}_r} \right) 
   + \frac{\tilde{b}_r\tilde{m}}{\pi\tilde{\rho}} \left( r\frac{d\ln\tilde{m}}{dr} 
  - D \frac{\tilde{b}_{\theta}}{\tilde{b}_r} \right) = 0,
 \label{eqn:EOMphi}
\end{equation}
again after substituting representative values for the remaining angular dependences. 

\section{From Static to Dynamic Equilibrium}

In considering the problem of jet formation, we have to carefully chose the starting point 
from which our chain of reasoning to begin. As such a point, we select the static balance 
of a gas sphere around the center of gravity, apart from the transversal equilibrium attained in a 
conical jet. When neglected the magnetic and inertial forces, equation (\ref{eqn:EOMr}) reduces to 
\begin{equation}
  \frac{1}{\tilde{\rho}}\ \frac{d\tilde{p}}{dr} + \frac{GM}{r^2} = 0, 
\end{equation}
and further with the aid of equation (\ref{eqn:gasp}), to 
\begin{equation}
   K\tilde{T}\ \frac{d\ln\tilde{p}}{dr} + \frac{GM}{r^2} = 0. 
\end{equation}

If the temperature varies with the radius according to a power law, 
 \begin{equation}
  \tilde{T}(z) = T_0 z^{-\alpha} \quad(\alpha\geq0), 
 \label{eqn:Txi}
 \end{equation}
then we can readily obtain the solution 
\begin{eqnarray}
 \lefteqn {\tilde{p}(z) } \nonumber \\
 & & = \left\{ \begin{array}{ll}
   \displaystyle{p_0\exp\left[-\frac{\kappa}{\alpha-1}\left\{ 
     z^{\alpha-1}-1\right\}\right],} &(\alpha>1) \\
   p_0 z^{-\kappa}, &(\alpha=1), \\
   \displaystyle{p_0\exp\left[-\frac{\kappa}{1-\alpha}\left\{
     1-z^{-(1-\alpha)}\right\}\right],} &(1>\alpha\geq0)
 \end{array} \right. 
 \label{eqn:stat}
\end{eqnarray}
where $\kappa\equiv GM/KT_0 r_{\rm J}$. 
It can be seen from this solution that the static equilibrium is realized (i.e., $\tilde{p}
\rightarrow 0$ as $z\rightarrow \infty$) only for $\alpha\geq 1$, and that the appearance 
of outflowing plasma is suggested for $1>\alpha\geq0$ since then $\tilde{p}$ remains finite 
even at $z\rightarrow \infty$. 

Extending this consideration to the case of a dynamical force balance realized in a jet under 
the presence of other forces than the gravity and pressure gradient, we assume that $\tilde{p}(z)$ 
takes a form like 
\begin{equation}
 \tilde{p}(z) = p_0 z^{-\beta}\exp\left[-\frac{\kappa}{1-\alpha}
   \left\{1-z^{-(1-\alpha)}\right\}\right], 
 \label{eqn:pxi}
\end{equation}
i.e., a mixture of marginal and outflow types in equation (\ref{eqn:stat}). Here, $\alpha$ 
and $\beta$ are both constants whose values are restricted in the ranges, $1>\alpha\geq0$ 
and $\beta\geq0$. For this choice, we can show easily that 
\begin{equation}
 -\frac{1}{\tilde{\rho}}\frac{d\tilde{p}}{dr} - \frac{GM}{r^2} = \beta\frac{K\tilde{T}}{r}. 
\end{equation}
This relation indicates that, when the parameter $\beta$ does not vanish, a part of the pressure 
gradient force survives after cancellation with the gravity. Speaking in other way, the functional 
form of the gravity remains in equation (\ref{eqn:EOMr}) through the form of the miss-cancelled 
pressure force. On the other hand, when $\beta$ exactly vanishes, nothing that can trace the gravity 
is left in equation (\ref{eqn:EOMr}), and only the MHD forces control the jets. 

Since the functional form of the pressure has been specified, it follows immediately from the 
equation of state (\ref{eqn:gasp}) that 
\begin{equation}
 \tilde{\rho}(z) = \rho_0 z^{\alpha-\beta}
   \exp\left[-\frac{\kappa}{1-\alpha}\left\{1-z^{-(1-\alpha)}\right\}\right], 
 \label{eqn:rho}
\end{equation}
where $\rho_0\equiv p_0/KT_0$, and further from equation (\ref{eqn:bpxi}) that 
\begin{eqnarray}
 \tilde{b}_r(z) &=& \tilde{b}_{\varphi}(z)  \nonumber \\
   &=& b_0 z^{-\beta/2}\exp\left[-\frac{\kappa}{2(1-\alpha)}
    \left\{1-z^{-(1-\alpha)}\right\}\right], 
 \label{eqn:bjet}
\end{eqnarray}
where $b_0 \equiv (8\pi p_0)^{1/2}$. Substituting the above expressions into equation 
(\ref{eqn:v2phi}), we have 
\begin{equation}
 \tilde{v}_{\varphi}(z) = (8KT_0)^{1/2}z^{-\alpha/2}. 
\end{equation}

The radial velocity of the jet is essentially controlled by the $\varphi$-component of Ohm's law. 
Substituting the expression for $\tilde{j}_{\varphi}$ from equation (\ref{eqn:tjphi}), we can 
rewrite equation(\ref{eqn:Ohphi}) in the from 
\begin{equation}
 \frac{\tilde{b}_{\theta}}{\tilde{b}_r} - \frac{\tilde{v}_{\theta}}{\tilde{v}_r} 
  + \frac{c^2}{2\pi\sigma\Theta^2r_{\rm J}v_{\infty}}\ \frac{v_{\infty}}{\tilde{v}_r}\ z^{-1} = 0. 
\end{equation}
Further inserting the results from the flux and mass conservations, we finally obtain the expression 
\begin{equation}
 \frac{d}{dz}\ln\left(\frac{\tilde{\rho}\tilde{v}_r}{\tilde{b}_r}\right) 
  = \Re_{\rm J}^{-1}\frac{v_{\infty}}{\tilde{v}_r}\ z^{-2}, 
 \label{eqn:asyV}
\end{equation}
where $\Re_{\rm J}\equiv 2\pi\sigma\Theta^2r_{\rm J}v_{\infty}/c^2$ is the magnetic Reynolds 
number characterizing the jet structures. We have to note here that the assumption of infinite 
conductivity ( $\sigma\rightarrow\infty$, i.e., the limit of ideal MHD) restricts the result 
obtained above too strictly (see also equation (\ref{eqn:vterm}) below).   

Among the asymptotic solutions of the above equation (\ref{eqn:asyV}), only the one that tends to 
a constant velocity $v_{\infty}$ at large distances (i.e., at $z\gg 1$) is of our interest here. 
Putting $\tilde{v}_r=v_{\infty}$ on the right hand side of equation (\ref{eqn:asyV}), we obtain 
by integration 
\begin{equation}
 \frac{(\tilde{\rho}/\rho_0)(\tilde{v}_r/v_0)}{\tilde{b}_r/b_0} 
  = \exp\left[\Re_{\rm J}^{-1}(1-z^{-1})\right]. 
\end{equation}
Further substituting expressions (\ref{eqn:rho}) and (\ref{eqn:bjet}), we have 
\begin{eqnarray}
 \lefteqn{ \tilde{v}_r(z) = v_0 z^{\beta/2-\alpha} } \nonumber \\
  & &\quad \times \exp \left[ \frac{\kappa}{2(1-\alpha)}\left\{1-z^{-(1-\alpha)}\right\} 
   + \Re_{\rm J}^{-1}(1-z^{-1}) \right]. 
\end{eqnarray}
In order to guarantee that this $\tilde{v}_r(z)$ actually tends to a constant value $v_{\infty}$ 
when $z\rightarrow\infty$, its power law dependence on $z$ should vanish. Here, we have to identify 
this condition very carefully. Since the power law dependence in an arbitrary function $F(z)$ 
causes a $z^{-1}$ term in the derivative $d\ln F/dz$, we discuss here 
\begin{equation}
  \frac{d}{dz}\ln\tilde{v}_r = \left(\frac{\beta}{2}-\alpha\right) z^{-1} 
   + \frac{\kappa}{2}\ z^{-(2-\alpha)} + \Re_{\rm J}^{-1}z^{-2}. 
\end{equation}

Within the allowed values for $\alpha$ ($0 \leq \alpha<1$), there are two possibilities as for whether 
the second term on the right-hand side belongs to the first or third term: 1) when $\alpha=1$ 
(or more exactly, $\alpha\rightarrow 1$) the vanishing of $z^{-1}$ term is attained if 
$\beta=2-\kappa$, and 2) when $\alpha=0$ the condition yields $\beta=0$. These cases may be called 
the ``virial-jet" and ``isothermal-jet" solutions, respectively, because of the behaviors in 
temperature. The detailed discussions of these solutions are given in the following two sections. 


%% file: pt2.tex

\section{Virial-Jet Solution}

\subsection{asymptotic behavior}

Although we have set $\alpha=1$ in the virial jet case, actually we need a more careful treatment 
of the argument in the exponential functions. Since $\alpha$ should be smaller than unity as far as 
an outflow exists at infinity, we have to write $\alpha=1-\epsilon$ ($\epsilon>0$) and take the 
limit of $\epsilon\rightarrow 0$. Then, the resulting expressions for the radial velocity is 
\begin{eqnarray}
 \tilde{v}_r(z) 
  &=& v_0 z^{-\kappa/2}\exp\left[\frac{\kappa}{2\epsilon}\left(1-z^{-\epsilon}\right) 
   + \Re^{-1}_{\rm J}\left(1-z^{-1}\right)\right]  \nonumber \\ 
  &=& v_{\infty}z^{-\kappa/2}\exp\left[ \frac{\kappa}{2\epsilon}
      \left(1-z^{-\epsilon}\right) - \Re^{-1}_{\rm J}z^{-1}\right], 
 \label{eqn:vrV}
\end{eqnarray}
where 
\begin{equation}
 v_{\infty}\equiv v_0\ \exp \left( \Re^{-1}_{\rm J} \right). 
 \label{eqn:vterm}
\end{equation}
Since $v_{\infty}$ represent the terminal velocity of a jet, the jet acceleration becomes the more 
effective the smaller the magnetic Reynolds number (or the conductivity) is. Especially, in the limit 
of infinite conductivity (i.e., $\Re_{\rm J}\rightarrow\infty$) no acceleration can be obtained. 
 
The apparent power law dependence, $z^{-\kappa/2}$, is actually cancelled by the exponential factor 
containing $\epsilon$, in the limit of $\epsilon\rightarrow 0$. This can be confirmed with the aid 
of the identity 
\begin{equation}
  \lim_{\epsilon\rightarrow 0} 
   \left[z^{-a}\exp\left\{\frac{a}{\epsilon}\left(1-z^{-\epsilon}\right)\right\}\right] = 1, 
 \label{eqn:IDT} 
\end{equation}
which holds for an arbitrary real number $a$. The constancy of the left-hand side of (\ref{eqn:IDT}) 
is guaranteed by the vanishing of its derivative with respect to $z$, and its value is fixed as 1 
by evaluating it at $z=1$. 

There is further subtlety to be considered, in this type of outflows. 
In the actual situations, every jet cannot extend to infinity but terminates at a large but finite 
radius owing to the interaction with the surrounding galactic or inter-galactic matter. It is 
therefore obvious that the jet solution too has the end point $r_{\rm out}$, beyond which it cannot 
be applied. In such non-ideal cases, $\epsilon$ can remain finite (i.e., the out flow at infinity 
does not exist) even if it may be very small. 
When this is the case, we cannot replace any exponential factors that contain $\epsilon$ by a power 
law expression, by using the identity (\ref{eqn:IDT}). However, as far as the jet is well developed, 
we may consider that $z_{\rm out}\equiv r_{\rm out}/r_{\rm J}\gg 1$ and the radial velocity already 
approaches a terminal velocity $v_{\infty}$ at $z=z_{\rm out}$. In this case, we have 
\begin{eqnarray}
  \lefteqn{ \exp\left[-\frac{\kappa}{2\epsilon}z_{\rm out}^{-\epsilon} 
   - \Re^{-1}_{\rm J}z_{\rm out}^{-1}\right] \simeq 1, } \nonumber \\  
  & & \quad  z_{\rm out}^{-\kappa/2}\exp\left(\frac{\kappa}{2\epsilon}\right) = 1,  
\end{eqnarray}
from equation (\ref{eqn:vrV}). The former relation means that the limit of $z_{\rm out}\rightarrow
\infty$ should be taken while keeping $\epsilon$ finite, and the latter equation yields 
\begin{equation}
  \epsilon = \left(\ln z_{\rm out}\right)^{-1}. 
\end{equation}

In order to discuss the behaviors in the asymptotic regions ($z\gg 1$) of the present kind of jets, 
it is convenient to introducing a new non-dimensional radius $\zeta\equiv r/r_{\rm out}=z/z_{\rm out}$ 
in which the top of a jet corresponds to $\zeta=1$. Then we obtain from equation (\ref{eqn:vrV}) 
\begin{eqnarray}
 \tilde{v}_r(z) 
  &=& v_{\infty} \left(\frac{z}{z_{\rm out}}\right)^{-\kappa/2}
   \exp\left[-\frac{\kappa}{2\epsilon}z^{-\epsilon}
   - \Re^{-1}_{\rm J}z^{-1}\right]  \nonumber \\
   &\longrightarrow& \ 
  v_{\infty}\ \zeta^{-\kappa/2} \equiv \tilde{v}_r(\zeta), 
 \label{eqn:vrAS}
\end{eqnarray}
where $\tilde{v}_r(\zeta)$ represents the asymptotic behavior of $\tilde{v}_r(z)$ in the limit of 
$z\rightarrow z_{\rm out}$. This implies that the actual jets (with only a finite length) of virial 
type experience rather deceleration in their asymptotic regions toward their terminal values 
$\tilde{v}_r(1)=v_{\infty}$. Such a behavior is strongly reminiscent of the jets in FR\ I radio 
galaxies (see section 1 and references therein). 

For other quantities of our interest, the original $z$-dependences and their asymptotic forms are 
summarized as follows: 
\begin{equation} 
 \tilde{T}(z)=T_0\ z^{-1} = T_{\infty}\ \zeta^{-1} \equiv \tilde{T}(\zeta), 
\end{equation}
\begin{eqnarray} 
 \tilde{p}(z) &=& p_0\ z^{-(2-\kappa)}
    \exp\left[-\frac{\kappa}{\epsilon}\left(1-z^{-\epsilon}\right)\right]  \nonumber \\
    &\longrightarrow& \ p_{\infty}\ \zeta^{-(2-\kappa)} \equiv \tilde{p}(\zeta), 
 \end{eqnarray}
\begin{eqnarray} 
 \tilde{\rho}(z) &=& \rho_0\ z^{-(1-\kappa)}
   \exp\left[-\frac{\kappa}{\epsilon}\left(1-z^{-\epsilon}\right)\right]  \nonumber \\
   &\longrightarrow& \ \rho_{\infty}\ \zeta^{-(1-\kappa)} \equiv \tilde{\rho}(\zeta), 
\end{eqnarray}
\begin{eqnarray} 
 \tilde{b}_r(z) &=& \tilde{b}_{\varphi}(z) = b_0\ z^{-(1-\kappa/2)}
   \exp\left[-\frac{\kappa}{2\epsilon}\left(1-z^{-\epsilon}\right)\right]  \nonumber \\ 
   &\longrightarrow& \ b_{\infty}\ \zeta^{-(1-\kappa/2)} 
  \equiv \tilde{b}_r(\zeta)=\tilde{b}_{\varphi}(\zeta), 
 \label{eqn:vrBp}
\end{eqnarray}
\begin{eqnarray}
 \tilde{v}_{\varphi}(z) &=& (8KT_0)^{1/2} z^{-1/2} \nonumber \\
  &=& (8KT_{\infty})^{1/2} \zeta^{-1/2} \equiv \tilde{v}_{\varphi}(\zeta), 
 \label{eqn:vpalp}
\end{eqnarray}
\begin{eqnarray}
 \tilde{j}_r(z) &=& j_0\ z^{-(2-\kappa/2)}
   \exp\left[-\frac{\kappa}{2\epsilon}\left(1-z^{-\epsilon}\right)\right]  \nonumber \\
   &\longrightarrow& \ j_{\infty}\ \zeta^{-(2-\kappa/2)} \equiv \tilde{j}_r(\zeta), 
\end{eqnarray}
\begin{eqnarray}
 \tilde{j}_{\varphi}(z) &=& 2j_0\ z^{-(2-\kappa/2)}
   \exp\left[-\frac{\kappa}{2\epsilon}\left(1-z^{-\epsilon}\right)\right]  \nonumber \\
   &\longrightarrow& \ 2j_{\infty}\ \zeta^{-(2-\kappa/2)} \equiv \tilde{j}_{\varphi}(\zeta), 
\end{eqnarray}
where $T_{\infty}=T_0/z_{\rm out}$, $p_{\infty}=p_0/z_{\rm out}^2$,  
$\rho_{\infty}=\rho_0/z_{\rm out}$, $b_{\infty}=b_0/z_{\rm out}$, 
$j_0\equiv cb_0/4\pi r_{\rm J}$, and $j_{\infty}=j_0/z_{\rm out}^2$. 
Since it is natural to expect that all the above quantities should decrease with radius $\zeta$, 
we obtain the restriction $0<\kappa<1$. 

In the asymptotic region of an actual jet of the virial type (i.e., $\epsilon\neq0$), 
the continuity equations for mass and magnetic flux yield 
\begin{equation}
  \tilde{v}_{\theta}(\zeta) 
   = v_{\infty} \left(1+\frac{\kappa}{2}\right)\zeta^{-(1-\kappa/2)}, 
\end{equation}
and 
\begin{equation}
  \tilde{b}_{\theta}(\zeta) 
   = b_{\infty} \left(1+\frac{\kappa}{2}\right)\zeta^{-\kappa/2}, 
\end{equation}
respectively, since $d\ln\tilde{v}_r/d\zeta=-(\kappa/2)\zeta^{-1}$ and 
$d\ln\tilde{b}_r/d\zeta=-(1-\kappa/2)\zeta^{-1}$. 
The $\theta$-component of the current density is obtained, from the relation (\ref{eqn:tjth}), as 
\begin{equation}
  \tilde{j}_{\theta}(\zeta)= \frac{\kappa}{2}\ j_{\infty}\ \zeta^{-(2-\kappa/2)}. 
\end{equation}
Within the jets of virial type, both sound and Alf\'en velocities, $C_{\rm S}$ and $V_{\rm A}$, 
vary as $\zeta^{-1/2}$: 
\begin{equation}
  C_{\rm S}^2 \equiv \frac{\tilde{p}}{\tilde{\rho}} = KT_0 z^{-1} = KT_{\infty} \zeta^{-1}, 
\end{equation}
\begin{equation}
  V_{\rm A}^2 \equiv \frac{\tilde{b}_r^2}{\pi\tilde{\rho}} 
    = \frac{\tilde{b}_{\varphi}^2}{\pi\tilde{\rho}} 
    = 8KT_0 z^{-1} = 8KT_{\infty} \zeta^{-1}. 
\end{equation}

\subsection{consistency}

Before the above obtained set of physical quantities are accepted for an actual solution to the full 
set of MHD equations, some more equations are remaining to be discussed. They are the $r$- and 
$\varphi$-components of EOM, and the stationarity condition expressed in terms 
of the electric field. These equations are checked below only in the asymptotic region. 

We begin with the $r$-component of EOM. In the asymptotic region, the $r$-dependence of 
the miss-cancelled pressure gradient force and the MHD-force are the same, i.e., 
$-\beta(KT/r)\propto r^{-2}$ and $\tilde{b}_{\varphi}^2/2\pi\tilde{\rho}r\propto r^{-2}$. 
This means that the MHD-force cannot dominate over the gravity even in the asymptotic region, and 
is consistent with the anticipation based on the idealized situation (see the previous section) that 
the out flow tends to the marginally bound case in the limit of $\alpha\rightarrow 1$. It turns out 
that the inertial terms dominate over other forces 
since $\tilde{v}_r^2/r\propto r^{-(1+\kappa)}$ and $1<1+\kappa<2$. 
Thus, the $r$-component of EMO reduces to the vanishing of these terms, 
\begin{eqnarray}
  \zeta\frac{d\ln\tilde{v}_r}{d\zeta} + 2A\frac{\tilde{v}_{\theta}}{\tilde{v}_r} = 0 
 \label{eqn:eomr}
\end{eqnarray}
within the accuracy to the leading order in $r$-dependences, which yields 
\begin{equation}
  A = \frac{\kappa}{2(2+\kappa)}. 
 \label{eqn:A}
\end{equation}
For a $\kappa$ in the range $0<\kappa<1$, we have $0<A<1/6$ which satisfies the general requirement 
$0<A<1/2$ discussed in section 3. 

Next, we check the $\varphi$-component (\ref{eqn:EOMphi}) of EOM. In the asymptotic region, 
$\tilde{l}$ and $\tilde{m}$ behave like $\tilde{l}\sim r^{1/2}$ and $\tilde{m}\sim r^{-\kappa/2}$, 
respectively. Therefore, the inertial forces behave like $\tilde{v}_r(d\tilde{l}/d\zeta) \sim 
\tilde{v}_{\theta}\tilde{l}/\zeta \sim r^{-(1-\kappa)/2}$, and the MHD forces, like
$(\tilde{b}_r/\pi\tilde{\rho})(d\tilde{m}/d\zeta) 
\sim \tilde{b}_{\theta}\tilde{m}/\pi\tilde{\rho}\zeta \sim r^{-1}$. 
In the leading order, this equation also requires the vanishing of the inertial terms, which 
reduces to 
\begin{equation}
  D = \frac{1}{\kappa+2}. 
 \label{eqn:D}
\end{equation}
For a $\kappa$ in the range $0<\kappa<1$, we have $1/3<D<1/2$ which is again consistent with 
the general requirement $0<D<1$. Thus, in this solution, the values of the parameters, $A$ and $D$, 
have been fixed consistently in terms of $\kappa$. 

The poloidal components of the electric field are calculated from the corresponding 
components of Ohm's law. 
The $\theta$-component is calculated from equation (\ref{eqn:Ohth}) as 
\begin{equation}
  \tilde{E}_{\theta}(\zeta) = -\frac{\tilde{b}_r\tilde{v}_r}{c}
   \left(1 + \frac{\tilde{v}_{\varphi}}{\tilde{v_r}}\right) 
  \simeq -\frac{b_{\infty}v_{\infty}}{c}\zeta^{-1}, 
 \label{eqn:vrEth}
\end{equation}
where the second term in the parentheses has been neglected since it is proportional to 
$r^{-(1-\kappa)/2}$. 
From equations (\ref{eqn:Ohr}) and (\ref{eqn:tjr}) we obtain for the $r$-component 
\begin{eqnarray}
  \tilde{E}_r(\zeta) &=& \frac{1}{c}\ \tilde{v}_r\tilde{b}_{\varphi} \left\{
    \frac{\tilde{v}_{\theta}}{\tilde{v}_r} + \frac{\tilde{v}_{\varphi}}{\tilde{v}_r}
    \ \frac{\tilde{b}_{\theta}}{\tilde{b}_r} 
    + \frac{c^2}{4\pi \sigma \Theta^2v_{\infty}}\frac{v_{\infty}}{\tilde{v}_r}z^{-1} \right\} 
  \nonumber \\ 
  &\simeq& \frac{1}{c}\left(1+\frac{\kappa}{2}\right) b_{\infty}v_{\infty}\ \zeta^{-1}, 
 \label{eqn:vrEr}
\end{eqnarray}
where the second and third terms in the curly brackets have been neglected because they are 
proportional to $r^{-(1-\kappa)/2}$ and $r^{-(1-\kappa/2)}$, respectively, while the first 
term remains finite. 

Substituting the above obtained results for $\tilde{E}_r$ and $\tilde{E}_{\theta}$ into equation 
(\ref{eqn:dbdt}), we have 
\begin{equation}
  \frac{\partial\tilde{b}_{\varphi}}{\partial t} 
   = -\left(1+\frac{\kappa}{2}\right)\frac{b_{\infty}v_{\infty}}{r_{\rm out}}\ \zeta^{-2}, 
 \label{eqn:nstb}
\end{equation}
which means that the strength of the toroidal magnetic field decreases with time, at least in 
the asymptotic region. This fact seems rather natural, however, because $b_{\infty}=b_0/z_{\rm out}$ 
decreases gradually with increasing $z_{\rm out}$ as the top of a jet proceeds into the ambient 
plasma. Even if this is the case, the characteristic time for this change is estimated to be very 
large: $t_{\rm mag} \sim \vert d\ln\tilde{b}_{\varphi}/dt \vert^{-1} \sim r_{\rm out}/v_{\infty}$, 
which gives $t_{\rm mag}\sim 10^7$ yr for $r_{\rm out}\sim100$ kpc and $v_{\infty}\sim10^9$ 
cm s$^{-1}$. Therefore, unless we are interested in the evolutional aspects of the jets, such 
a small change is negligible as usually done in the current jet theories. 

\subsection{global aspects}

In relation to the virial-jet solution, it should be emphasized that the solution can be reconciled 
naturally with RIAF (i.e., radiatively-inefficient accretion flow)-type accretion disks in the 
external magnetic field \citep{Kab00,Kab01,Kab07}. This type of accretion model has a virial-type 
high temperature and too tenuous plasma density to be able to radiate efficiently. 
In contrast to the familiar RIAF models (e.g., \cite{NY94,Ab95,NY95}), however, 
this model explicitly takes into account the presence of an ordered magnetic field in the galactic 
nuclear regions. Therefore, the extraction of angular momentum from the accretion flow is governed 
by the magnetic stress caused by the twisted magnetic field, instead of the viscous stress. 
It has also been clarified that a transport of heat energy (i.e., a non-adiabaticity) through the 
non-radiating disc causes a wind from the disc surface \citep{Kab01,MKab03}, but this wind should 
be distinguished from jets that emanate from the inner edges of such accretion discs.

Another feature to be mentioned is that the disc has a geometrically thin structure in spite of its 
high temperature. This is because the disc plasma is vertically compressed toward the equatorial 
plane by the magnetic pressure due to the toroidal field developed outside (i.e., above and below) 
the disc. The addition of this component is due to the twisting-up of the seed field by a rotational 
motion of the accreting plasma, and the degree of which is determined by the balance between the 
twisting motion and the slip-out (or magnetic diffusion) of the poloidal field under the presence 
of a finite electrical resistivity. 
In terms of the circuit theory, the origin of this twist can be attributed to the existence of 
a poloidally circulating electric current. As already mentioned in section 2, such a current is 
actually driven by the accretion disc operating as a current generator and should close finally 
through the polar jet regions. 

The $\varphi$-component of the magnetic filed developed above and blow an RIAF-type accretion disc 
is given \citep{Kab01} by 
\begin{equation}
  \tilde{b}_{\varphi}(\xi) = \Re_{\rm D}\vert B_0 \vert\ \xi^{-(1-n)}, 
\end{equation}
where $\xi\equiv r/r_{\rm A}$ is the radius normalized by the size of disc's outer edge, $r_{\rm A}$. 
Further, $\Re_{\rm D}$ is the magnetic Reynolds number characterizing the accretion disc, and 
the parameter $n$ specifies the strength of winds, which is a measure of the strength of the 
heat transport (or the non-adiabaticity) in the non-radiating disc plasma. Comparing the above 
expression with the asymptotic form $\tilde{b}_{\varphi}(\zeta)$ in (\ref{eqn:vrBp}), we realize 
that these expressions coincide strictly under the identification of 
\begin{equation}
   n=\frac{\kappa}{2}, 
\end{equation} 
except for the difference in the normalizing factors for the radius. 
Therefore, the inequality $0<\kappa<1$ reduces to $0<n<1/2$, which means that there is no downward 
wind ($n<0$) as far as an accretion disc is driving a jet. 
Substituting the expression for $\kappa$, we can specify the location of the footpoint of a jet 
in terms of the wind parameter as 
\begin{equation}
  r_{\rm J} = \frac{GM}{2nKT_0} > \frac{GM}{KT_0}. 
 \label{eqn:foot}
\end{equation}
Since $r_{\rm J}\rightarrow\infty$ as $n\rightarrow0$, a nearly adiabatic accretion disc 
($n\simeq 0$) seems to be able to drive only a weak, and therefore undetectable, jet. 

The coincidence of the toroidal magnetic fields generated by an accretion disc and by a jet implies 
that the origin of this field is one and the same current circulating through both disc and jet. 
Equating the field strength of both solutions at $r=r_{\rm A}$ (i.e., $\xi=1$, 
or $\zeta=\zeta_{\rm A}\equiv r_{\rm A}/r_{\rm out}<1$), we obtain 
\begin{equation}
  b_0=\Re_{\rm D}z_{\rm out}\zeta_{\rm A}^{1-n}\ \vert B_0 \vert 
     =\Re_{\rm D}z_{\rm A}\zeta_{\rm A}^{-n}\ \vert B_0 \vert,
\end{equation}
which depends on $r_{\rm out}$ unless $n=0$. This specifies the central field strength $b_0$ 
in terms of the external field $\vert B_0 \vert$. From this result, we can confirm that 
$b_{\infty}=b_0/z_{\rm out}=\Re_{\rm D}\zeta_{\rm A}^{1-n}\ \vert B_0\vert$  decreases 
with increasing $r_{\rm out}$ since $1-n>0$. 

For one side of a pair of jets, mass ejection rate and integrated kinetic energy flux in the  
asymptotic region are calculated, respectively, as 
\begin{eqnarray}
  \dot{M}_{\rm J}(\zeta) &\equiv& \int_0^{\pi/2} (\rho v_r)\ 2\pi r^2 
   \sin\theta d\theta  \nonumber \\
  &=& \frac{4\pi}{27}\ \Theta^2 r_{\rm J}^2\rho_0v_{\infty}z_{\rm out}
   \ \zeta^{1+\kappa/2}, 
\end{eqnarray}
\begin{eqnarray}
  F(\zeta) &\equiv& \int_0^{\pi/2} \left(\frac{1}{2}\rho v^2\right) 
     2\pi r^2 \sin\theta d\theta  \nonumber \\
  &\simeq& \frac{\pi}{20}\Theta^2r_{\rm J}^2\rho_0 v_{\infty}^3z_{\rm out}\ \zeta^{1-\kappa/2}. 
\end{eqnarray}
Although the range of integration on $\theta$ is extended to its full range ($0\leq\theta\leq\pi/2$), 
for simplicity, this may be allowed since the matter density falls off rapidly to zero outside 
the jet cone. 
These quantities are both increasing functions of $\zeta$, reflecting that the velocity field has 
a converging component toward the polar axis (i.e., $v_{\theta}<0$). Of course, however, these 
values remain finite with their terminal values at the top of a jet being $\dot{M}_{\rm J}(1)$ 
and $F(1)$. 

Three components of the Poynting flux in the asymptotic region are 
\begin{eqnarray}
  \lefteqn{ P_r(\zeta, \eta)
   =\tilde{P}_r(\zeta)\ \mbox{sech}^2\eta\tanh^2\eta,} \nonumber \\
  & & \qquad \tilde{P}_r(\zeta) = -\frac{c}{4\pi}\tilde{E}_{\theta}\tilde{b}_{\varphi} 
   =\frac{b_{\infty}^2v_{\infty}}{4\pi}\ \zeta^{-(2-\kappa/2)}, 
\end{eqnarray}
\begin{eqnarray}
  \lefteqn{ P_{\theta}(\zeta, \eta)
   = -\Theta\tilde{P}_{\theta}(\zeta)\ \mbox{sech}^2\eta\tanh\eta,} \nonumber \\
  & & \qquad \tilde{P}_{\theta}(\zeta) = \frac{c}{4\pi}\tilde{E}_r\tilde{b}_{\varphi} 
   = \frac{b_{\infty}^2v_{\infty}}{4\pi} \left(1+\frac{\kappa}{2}\right) \zeta^{-(2-\kappa/2)}, 
\end{eqnarray}
\begin{eqnarray}
  \lefteqn{ P_{\varphi}(\zeta, \eta)
   =\tilde{P}_{\varphi}(\zeta)\ \mbox{sech}^4\eta\tanh\eta, } \nonumber \\
  & & \qquad \tilde{P}_{\varphi}(\zeta) = -\frac{c}{4\pi}\tilde{E}_{\theta}\tilde{b}_r 
   = \frac{b_{\infty}^2v_{\infty}}{4\pi}\ \zeta^{-(2-\kappa/2)}. 
\end{eqnarray}
The radial flow of electromagnetic energy through one side of a jet pair is defined as 
\begin{equation}
  S(\zeta)\equiv\int_0^{\pi}2\pi r^2 P_r \sin\theta d\theta 
   = \frac{1}{6}\Theta^2 r_{\rm J}^2b_0^2v_{\infty}\ \zeta^{\kappa/2}. 
\end{equation}
This quantity is again an increasing function of $\zeta$, reflecting the convergence of the magnetic 
field (i.e., $b_{\theta}<0$). 

The ratio of the Poynting flux to the kinetic flux, 
\begin{equation}
  \frac{S(\zeta)}{F(\zeta)} = \frac{10}{3}\ \frac{8KT_0}{v_{\infty}^2z_{\rm out}}\ \zeta^{-(1-\kappa)}, 
\end{equation}
is a decreasing function of $\zeta$, and its terminal value is 
$S(1)/F(1)=(80/3)(KT_0/v_{\infty}^2z_{\rm out})$. If we assume roughly that the dominance of Poynting 
flux should be kept in the virial jets all the way to their end points, the terminal velocity can be 
estimated as 
\begin{equation}
  v_{\infty}^2 = \frac{80}{3}\ \frac{KT_0}{z_{\rm out}}, 
\end{equation}
by setting $S(1)/F(1)\sim1$. In this case, the terminal velocity decreases with the jet length 
according to $z_{\rm out}^{-1/2}$, which also reflects the marginally trapped nature of the jets of 
this type. 

\section{Isothermal-Jet Solution}

\subsection{consistency}

The isothermal-jet solution is specified by the set of parameters, $\alpha=0$ and $\beta=0$. 
In this case, the relevant quantities take the following forms: 
\begin{equation}
  \tilde{T}(z) = T_0 = \mbox{const.}, 
\end{equation}
\begin{equation}
  \tilde{p}(z) = p_0\exp\left\{-\kappa\left(1-z^{-1}\right)\right\} 
   = p_{\infty}\exp\left(\kappa z^{-1}\right), 
\end{equation}
\begin{equation}
  \tilde{\rho}(z) = \rho_0\exp\left\{-\kappa\left(1-z^{-1}\right)\right\}
   = \rho_{\infty}\exp\left(\kappa z^{-1}\right), 
\end{equation}
\begin{eqnarray}
  \tilde{b}_r(z) &=& \tilde{b}_{\theta}(z)  \nonumber \\
   &=& b_0\exp\left\{-\frac{\kappa}{2}\left(1-z^{-1}\right)\right\}
   = b_{\infty}\exp\left(\frac{\kappa}{2} z^{-1}\right), 
\end{eqnarray}
\begin{eqnarray}
  \tilde{v}_r(z) 
  &=& v_0\exp\left\{\left(\Re_{\rm J}^{-1}+\frac{\kappa}{2}\right)\left(1-z^{-1}\right)\right\} 
  \nonumber \\
   &=& v_{\infty}\exp\left\{-\left(\Re_{\rm J}^{-1}+\frac{\kappa}{2}\right)z^{-1}\right\}, 
\end{eqnarray}
\begin{equation}
  \tilde{v}_{\varphi}(z) = (8KT_0)^{1/2} = \mbox{const.}, 
\end{equation}
where $p_{\infty}=p_0\exp(-\kappa)$, $\rho_{\infty}=\rho_0\exp(-\kappa)$, 
$b_{\infty}=b_0\exp(-\kappa/2)$ and $v_{\infty}=v_0\exp(\kappa/2+\Re^{-1})$. 

Since 
\begin{eqnarray}
  \lefteqn{ \frac{d}{dz}\ln\tilde{b}_r = \frac{d}{dz}\ln\tilde{b}_{\varphi} 
   = -\frac{\kappa}{2}\ z^{-2}, } \nonumber \\
  & & \qquad \frac{d}{dz}\ln\left(z^2\tilde{b}_r\right) 
   = \left(2-\frac{\kappa}{2}\right) z^{-1}, 
\end{eqnarray}
\begin{eqnarray}
  \lefteqn{ \frac{d}{dz}\ln\tilde{v}_r = \left(\frac{\kappa}{2}+\Re^{-1}\right) z^{-2}, } 
  \nonumber \\
  & & \qquad \frac{d}{dz}\ln\left(z^2\tilde{\rho}\tilde{v}_r\right) 
  = \left\{2-\left(\frac{\kappa}{2}-\Re^{-1}\right) z^{-1}\right\} z^{-1}, 
\end{eqnarray}
we obtain 
\begin{eqnarray}
  \tilde{b}_{\theta}(z) &=& b_0\left(2-\frac{\kappa}{2} z^{-1}\right)
   \exp\left\{-\frac{\kappa}{2}\left(1-z^{-1}\right)\right\} \nonumber \\
   &=& b_{\infty}\left(2-\frac{\kappa}{2} z^{-1}\right) \exp\left(\frac{\kappa}{2}z^{-1}\right)
\end{eqnarray}
and 
\begin{eqnarray}
  \tilde{v}_{\theta}(z) 
   &=& v_0\left\{2-\left(\frac{\kappa}{2}-\Re^{-1}\right) z^{-1}\right\} \nonumber \\
  & & \quad \times
   \exp\left[\left(\Re_{\rm J}^{-1}+\frac{\kappa}{2}\right)\left(1-z^{-1}\right)\right] 
  \nonumber \\ 
   &=& v_{\infty}\left\{2-\left(\frac{\kappa}{2}-\Re^{-1}\right) z^{-1}\right\} \nonumber \\
   & & \quad\times\exp\left[-\left(\Re_{\rm J}^{-1}+\frac{\kappa}{2}\right)z^{-1}\right]. 
\end{eqnarray}
Further, for the characteristic velocities, we have 
\begin{equation}
  C_{\rm S}^2 \equiv \frac{\tilde{p}}{\tilde{\rho}} = KT_0 = \mbox{const.}, 
\end{equation}
\begin{equation}
  V_{\rm A}^2 \equiv \frac{\tilde{b}_r^2}{\pi\tilde{\rho}} 
    = \frac{\tilde{b}_{\varphi}^2}{\pi\tilde{\rho}} 
    = 8KT_0 = \mbox{const}. 
 \label{eqn:vAlf}
\end{equation}
In contrast to the virial-jet solution, the isothermal-jet solution has an essentially exponential 
structure, and the quantities remain finite even at large distances ($z\rightarrow\infty$). 

As in the case of the viral-jet solution, we have next to check the consistency of the isothermal-jet 
solution. In the $r$-component of EOM, the inertial term tends to $Av_{\infty}^2$ 
since $\tilde{v}_r\rightarrow v_{\infty}$ as $z\rightarrow\infty$, while the MHD-force term tends 
to $-8KT_0D$. Therefore, in the leading order, the equation reduces to 
\begin{equation}
  v_{\infty}^2 A - 8KT_0 D = 0. 
 \label{eqn:vinf2}
\end{equation}
As for the $\varphi$-component of EOM, we find $\tilde{l} \equiv r\tilde{v}_{\varphi}\propto r$ and 
$\tilde{m}\equiv r\tilde{b}_{\varphi}\propto r$ in the asymptotic region ($z\gg1$). Since the inertial 
forces and MHD forces become of the same orders of magnitude, i.e., 
$(\tilde{b}_r/\pi\tilde{\rho})(d\tilde{m}/dz) \sim \tilde{b}_{\theta}\tilde{m}/\pi
\tilde{\rho}z \sim z^0$ and $\tilde{v}_r(d\tilde{l}/dz) \sim \tilde{v}_{\theta}
\tilde{l}/z \sim z^0$, equation (\ref{eqn:EOMphi}) is held only when 
\begin{equation}
  z\frac{d\ln\tilde{l}}{dz} - D\ \frac{\tilde{v}_{\theta}}{\tilde{v}_r} 
  = z\frac{d\ln\tilde{m}}{dz} - D\ \frac{\tilde{b}_{\theta}}{\tilde{b}_r} = 0, 
 \label{eqn:eompisoj}
\end{equation}
and both of them are satisfied if  
\begin{equation}
  D = \frac{1}{2}. 
 \label{eqn:D}
\end{equation}
Further, if the approximation 
\begin{equation}
  D = \left\langle \frac{f(\eta)}{\tanh\eta} \right\rangle 
    \simeq \frac{\langle f(\eta)\tanh\eta\rangle}{\langle\tanh^2\eta\rangle} = 2A  
\end{equation} 
is allowed, we can fix as $A=1/4$ and $D=1/2$, and the terminal velocity of a jet of this type is 
specified as 
\begin{equation}
  v_{\infty} = 4(KT_0)^{1/2}. 
\end{equation}
Speaking more generally, it can be concluded from equations (\ref{eqn:vinf2}) and (\ref{eqn:vAlf}) 
that the flow becomes super-Alfv\'enic as far as $D>A$. 

The poloidal components of the electric field are obtained from Ohm's law. In the asymptotic region 
($z\gg1$), they are 
\begin{equation}
  \tilde{E}_{\theta} = -\frac{\tilde{b}_r\tilde{v}_r}{c}
   \left(1+\frac{\tilde{v}_{\varphi}}{\tilde{v}_r}\right)
  = -\frac{b_{\infty}v_{\infty}}{c}\left(1+\sqrt{\frac{A}{D}}\right) 
\end{equation} 
and  
\begin{eqnarray}
  \tilde{E}_r(z) &=& \frac{1}{c}\ \tilde{b}_r\tilde{v}_r\left\{
   \frac{\tilde{v}_{\theta}}{\tilde{v}_r}+\frac{\tilde{v}_{\varphi}}{\tilde{v}_r} 
   \ \frac{\tilde{b}_{\theta}}{\tilde{b}_r} + \frac{c^2}{4\pi\sigma\Theta^2v_{\infty}}
  \frac{v_{\infty}}{\tilde{v}_r} z^{-1}\right\} \nonumber \\
   &\simeq& \frac{2}{c}\ b_{\infty}v_{\infty}\left(1+\sqrt{\frac{A}{D}}\right), 
 \label{eqn:isEr}
\end{eqnarray}
where the third term in the curly brackets has been neglected. 

As in the case of the viral jets, these field components do not satisfy the stationarity condition. 
The associated change rate of the magnetic field is given by 
\begin{equation}
  \frac{\partial\tilde{b}_{\varphi}}{\partial t} 
   = -\frac{2}{5}\frac{b_{\infty}v_{\infty}}{r_{\rm J}}\left(1+\sqrt{\frac{A}{D}}\right)\ z^{-1}. 
\end{equation}
The characteristic time for the changing magnetic field is given by 
$t_{\rm mag}\sim r_{\rm out}/v_{\infty}$, which is also large as in the case of the virial jets. 
In both cases, the long time scales can be understood as a consequence of large inductances of the 
relevant current systems \citep{Ben06}. In the equations above and in the the next subsection, 
we may substitute the value in (\ref{eqn:D}) for $D$. 

\subsection{global aspects}

Since we do not know any analytic solution of the accretion disk in a large-scale magnetic field, 
which is naturally connected to the isothermal-jet solution, we cannot specify 
the strength of the magnetic field $b_0$, in relation to the accretion process. 
The mass and kinetic energy flux through one side of a jet are calculated in the asymptotic 
region as 
\begin{equation}
  \dot{M}_{\rm J}(z) = \frac{4\pi}{27}\ \Theta^2 r_{\rm J}^2\rho_{\infty}v_{\infty}z^2 , 
\end{equation}
\begin{equation}
  F(z) = \frac{\pi}{20}\Theta^2r_{\rm J}^2\rho_{\infty}v_{\infty}^3z^2. 
\end{equation}
These are increasing functions of $z$, since the jet flow is again a converging one. Although these 
quantities diverge as $z\rightarrow\infty$, practically there is no jet that extends to infinity. 
If we want to know the exact nature of a jet expanding into real vacuum, then we must trace its 
entire evolution. However, such a problem is beyond the scope of the present paper. 

The three components of the Poynting flux in the asymptotic region are 
\begin{equation}
   \tilde{P}_r(z) = -\frac{c}{4\pi}\tilde{E}_{\theta}\tilde{b}_{\varphi} 
   =\frac{b_{\infty}^2v_{\infty}}{4\pi}\left(1+\sqrt{\frac{A}{D}}\right), 
\end{equation}
\begin{equation}
  \tilde{P}_{\theta}(z) = \frac{c}{4\pi}\tilde{E}_r\tilde{b}_{\varphi} 
   = \frac{b_{\infty}^2v_{\infty}}{2\pi}\left(1+\sqrt{\frac{A}{D}}\right), 
\end{equation}
\begin{equation}
  \tilde{P}_{\varphi}(z) = -\frac{c}{4\pi}\tilde{E}_{\theta}\tilde{b}_r 
   = \frac{b_{\infty}^2v_{\infty}}{4\pi}\left(1+\sqrt{\frac{A}{D}}\right). 
\end{equation}
Then, the integrated flux becomes 
\begin{equation}
  S_r(z) = \frac{1}{6}\left(1+\sqrt{\frac{A}{D}}\right)
    \Theta^2 r_{\rm J}^2b_{\infty}^2v_{\infty}z^2. 
\end{equation}
This is also an increasing function of $z$, reflecting the convergence of the magnetic field 
(i.e., $b_{\theta}<0$). 
The ratio of the Poynting flux to the kinetic flux is given by 
\begin{equation}
  \frac{S(z)}{F(z)} = \frac{10}{3}\left(1+\sqrt{\frac{A}{D}}\right)\frac{A}{D}, 
\end{equation}
which is independent of $z$. Thus, the isothermal jets seem to remain always Poynting-flux dominated. 

\section{Summary and Discussion}

We have found two types of jet solutions to the set of MHD equations, which are valid in the region 
except very close to the footpoint, under the restriction that the jet plasma is confined within 
a thin sheath structure centered on a conical surface of small opening angle. Commonly to both types 
of jet, such a density enhancement of jet plasma (assumed to be a normal plasma consisting of electrons 
and ions) is maintained in the valley of the magnetic pressures that is formed between the region of 
poloidal-component dominance near the polar axis and that of azimuthal-component dominance outside 
the cone. In this configuration, further, the centrifugal force due to jet rotation is always balanced 
by the hoop stress of the toroidal magnetic field. Although we have not discussed explicitly for it, 
the spine region is very likely to be filled with highly relativistic electron-positron pair plasma 
as various observations seem to suggest. The terminal velocity of a jet of both types of solution 
does not exceed by far the thermal velocity at its footpoint, and so remains to be at most mildly 
relativistic.  

The virial-jet solution is characterized by a virial-like temperature profile, and it seems that 
the jets are marginally bound by the gravity. Namely, although the radial velocity of a jet increases 
toward a terminal velocity at infinity in the case of free expansion, when the jet terminates 
at a large but finite distance, as is the case in every actual situations, the velocity decreases 
according to a power law in the asymptotic region to reach a terminal value. All other quantities 
also vary according to power laws. This type of solution can be matched quite naturally with the RIAF 
type solution of accretion discs in a large-scale magnetic field \citep{Kab00,Kab01,Kab07}. This fact 
strongly suggests that the virial jets are driven in the actual situations by such accretion dicks. 
The ratio of the Poynting flux to kinetic flux decreases with the radius, and may 
decrease to a value smaller than unity. 

On the other hand, the isothermal-jet solution is characterized by a temperature independent of the 
radius, and the jets are accelerated purely magnetohydrodynamically, since the gravity is exactly 
cancelled out by the pressure gradient force. The radial velocity is gradually accelerated and 
finally seems to reach a super-Alfv\'enic value. In this solution, however, the quantities other 
than the temperature vary as exponentials of inverse radius, and the changes in their values from 
the footpoint to the top remain rather small. The ratio of Poynting to kinetic flux is a constant 
in radius and remains always Poynting-flux dominated. Further, there is no accretion disk solution 
hitherto known to us, which can be matched naturally with this type of jet solution. 

The correspondence of the virial jets to FR\ I jets may be already evident from the above description. 
The FR\ I jets are known from observations to have well defined opening angles and decreasing radial 
velocities at large distances. The latter fact is explained here by their marginally bound nature. 
The opinion that the matter entrainment is the cause of deceleration (as often suggested in the 
literature, e.g., \cite{Bick94}) is not confirmed in our solutions. Although the virial-jet solution 
indeed has the flow and magnetic field that converge toward the jet axis, and hence the so-called 
entrainment of surrounding matter is realized there, such a configuration is also found in our 
isothermal-jet solution in which the radial velocity does not experience a deceleration. 

The host of FR\ I jets are ellipticals often in rich cluster environments, and their accretion 
discs are believed to be radiatively inefficient. The smooth matching of the virial-jet solution 
to the RIAF-type accretion disc solution seems to be a strong support of the idea that a RIAF 
in a global magnetic field always drive a virial jet. The strength of such a jet is very likely 
to be controlled by the parameter $\kappa$, which specifies the height of jet's footpoint. Since 
this value has been determined as twice of the wind parameter $n$ (defined in \cite{Kab01}), which 
is a measure of non-adiabaticity in the accreting matter, the strength of a virial jet may also 
be controlled by the strength of such a radial energy transport taking place in the accretion disc. 

Although the connection of the isothermal-jet solution to FR\ II jets may be not so clear as the 
connection of the virial-jet solution to FR\ I jets, it seems nevertheless to be possible. The 
overall trend that the radial changes of various quantities are rather mild in this solution seems 
to be favorable to FR\ II jets. In contrast to the virial-jet solution, the jet velocity in this 
solution increases gradually until the end point is reached, in spite of the presence of an 
entrainment (i.e., a convergent flow). 

Finally, we refer to a rather speculative possibility of interpreting FR\ II jets, i.e., the idea 
that FR\ II jets are dying systems, in which their central engines have been switched off. 
Originally, this idea was proposed with a support found in the trend seen in the power-size (P-D) 
diagram of FR\ II sources, which seems to suggest an decrease in the jet luminosities with their 
ages (\cite{Bld82}). If this is the case, a jet that has been driven once as a RF\ I jet by 
the gas supply owing to some past event of galaxy encounter is now switched off, for the lack of 
matter supply, and is dying as a FR\ II jet. Also, we may see recurrences of the jet phenomena 
in rich clusters, because there may be periods of inefficient matter supply between successive 
galaxy encounters. In fact, examples of such recurrences can be seen in some examples as the 
successive bubbles seen in X-rays (for M87, see e.g., \cite{For07}). 

Other kind of radio sources, such as compact symmetric objects (or CSO, see e.g., \cite{Rea96}) 
that resemble FR\ II sources but are much smaller and hence believed to be young, have also been 
interpreted as dying systems. The problem of over population that may arise when we assume an 
evolutionary track from CSOs into FR\ II sources can thus be avoided. 
Most of the young sources will die young as CSOs (e.g., \cite{Gir07}). Only a small fraction 
of young systems can evolve as FR\ I sources to a big size, and then die as well-developed FR\ II 
sources when their central engines (or more properly speaking, dynamos) are switched off. Therefore, 
large FR\ II sources such as Cyg A seem to be very rare cases. 

If we take this interpretation of FR\ II sources seriously, it becomes likely that the morphology 
of their jet cavities, which are revealed by recent X-ray observations, are different from those 
of FR\ I sources. Jet cavities of RF\ I sources seem to be typically of hour-glass type that have 
narrow necks near the equatorial plane. This fact can easily been understood as reflecting 
the presence of infalling matter associated with their accretion discs (see figure 1a). 
On the other hand, the shape of the jet cavity of a switched off accretion system is likely to have 
a spindle shape with no neck in the equatorial plane (see figure 1b). This is understood as follows. 
The poloidal current flowing through a switched-off jet is maintained by the self-inductance 
of the system. The current flowing in the jet closes its circuit through the equatorial plasma disc 
and through the walls of the X-ray cavity in the northern and southern hemispheres. Different 
from the accretion-driven cases, however, the inductive electric field appears in the equatorial disc 
having the same sign as the electric current, trying to keep the current constant. Since the magnitude 
of the electric current cannot increase in spite of this inductive field, the rotational motion of 
the disc is forced to increase in order to satisfy Ohm's law, equation (\ref{eqn:Ohr}). The 
unbalanced centrifugal force caused by this super-Keplerian rotation drives the disc plasma outward, 
resulting in a secretion disc (an inverse of the accretion disc). 

In this respect, it is interesting to note that Cyg A, a typical of FR\ II sources, seems to have 
a spindle shaped X-ray cavity \citep{WSY06} although further observations are needed for confirmation. 
The equatorial rings of hot gasses seen in their X-ray images might be reflecting the existence 
of outflows. Anyway, if the above quoted interpretation of FR\ II jets is correct, similar spindle 
shaped X-ray cavities are expected also for other FR\ II sources.
